\documentclass[preprintnumbers, floatfix, preprintnumbers, letterpaper, superscriptaddress,nofootinbib, onecolumn]{revtex4}
\pdfoutput=1
\usepackage{graphicx}
\usepackage{microtype}
\usepackage{amsmath}
\usepackage{amssymb}
\usepackage{subfigure}
\usepackage{hyperref}
\usepackage{url}
\usepackage{xcolor}
\usepackage{color}
\usepackage{mathrsfs}
\usepackage{calrsfs}
\usepackage{amsfonts}
\usepackage{latexsym}
\usepackage{ragged2e}
\usepackage{epsfig}
\usepackage{textcomp}
\usepackage{eufrak}
\usepackage{tabularx}
\usepackage{eucal}
\usepackage{epsfig}
\usepackage{amsmath} 
\usepackage[utf8]{inputenc}
\usepackage{fourier} 
\usepackage{array}
\usepackage{makecell}
\usepackage{longtable}

\definecolor{vividviolet}{rgb}{0.62, 0.0, 1.0}
\definecolor{amaranth}{rgb}{0.9, 0.17, 0.31}
\definecolor{palatinateblue}{rgb}{0.15, 0.23, 0.89}
\definecolor{brightpink}{rgb}{1.0, 0.0, 0.5}
\definecolor{cornflowerblue}{rgb}{0.39, 0.58, 0.93}
\definecolor{deepcarminepink}{rgb}{0.94, 0.19, 0.22}
\definecolor{radicalred}{rgb}{1.0, 0.21, 0.37}

\hypersetup{ linktoc=all,
    colorlinks, linkcolor={palatinateblue},
    citecolor={brightpink}, urlcolor={amaranth}
}

\graphicspath{{Images/}}

\renewcommand{\d}[1]{\ensuremath{\operatorname{d}\!{#1}}}

\def\sideremark#1{\ifvmode\leavevmode\fi\vadjust{\vbox to0pt{\vss
 \hbox to 0pt{\hskip\hsize\hskip1em
 \vbox{\hsize1.3cm\tiny\raggedright\pretolerance10000
 \noindent #1\hfill}\hss}\vbox to8pt{\vfil}\vss}}}%
\def\beq{\begin{equation}}
\def\eeq{\end{equation}}

\begin{document}

\title{Holographic Principle and the Second Law in Stephani Cosmology Revisited}

\author{Daniele \surname{Gregoris}}
\email{danielegregoris@libero.it}
\affiliation{Center for Gravitation and Cosmology, College of Physical Science and Technology, Yangzhou University, \\180 Siwangting Road, Yangzhou City, Jiangsu Province  225002, China}
\affiliation{School of Aeronautics and Astronautics, Shanghai Jiao Tong University, Shanghai 200240, China}

\author{Yen Chin \surname{Ong}}
\email{ycong@yzu.edu.cn}
\affiliation{Center for Gravitation and Cosmology, College of Physical Science and Technology, Yangzhou University, \\180 Siwangting Road, Yangzhou City, Jiangsu Province  225002, China}
\affiliation{School of Aeronautics and Astronautics, Shanghai Jiao Tong University, Shanghai 200240, China}

\author{Bin \surname{Wang}}
\email{wangb@yzu.edu.cn}
\affiliation{Center for Gravitation and Cosmology, College of Physical Science and Technology, Yangzhou University, \\180 Siwangting Road, Yangzhou City, Jiangsu Province  225002, China}
\affiliation{School of Aeronautics and Astronautics, Shanghai Jiao Tong University, Shanghai 200240, China}

\begin{abstract}
We show in a model-independent way that the inhomogeneous cosmological class II Stephani model fulfills both the  cosmological holographic principle, and that the entropy is increasing with time. By this we mean the result does not depend on any assumption on the time evolution of the scale factor, or on the matter content of the Universe, we also do not need to use the numerical values of the cosmological parameters, which are inferred in the framework of the usual homogeneous Friedmann model. Therefore our analysis is not affected by the tension of the current estimates of the Hubble parameter, and does not rely on any model-dependent assumption of the entropy amount at the present epoch. Our analysis allows us to set an upper limit for the inhomogeneity parameter of the model, an upper bound for the size that this type of universe can reach during the time evolution, a lower bound for the entropy abundance, and an estimate of the present day value of the deceleration parameter.
\end{abstract}

\maketitle

\section{Introduction: Stephani Cosmology and Its Problems}

According to the central tenet of general relativity, gravity is the manifestation of spacetime curvature. What then is the geometry of our actual Universe\footnote{We use lower case ``universe'' to describe a generic, hypothetical spacetime, and uppercase ``Universe'' for the one we actually living in.}?
The search for the best cosmological model can be roughly divided into two distinct phases: the theoretical model building and the test against available astrophysical datasets. The former can be addressed by choosing a certain gravitational theory (general relativity is the most common adopted nowadays, but extensions beyond it have been investigated as well), derive the field equations through a variational principle, and solve them based on some assumptions about the large scale geometry of the Universe and its matter content. When the differential equations at the core of the theory are solved, some free integration parameters appear, and they can be constrained by considering the data provided by astrophysical observations. Of course, it is possible that no matter how one fixes the numerical values of such free parameters the astrophysical datasets cannot be accounted for under a chosen model. The most pedagogical example that can illustrate this point is the static Einstein universe, which cannot account for the Hubble law describing the cosmic flow of the galaxies from each other regardless the amount of matter-energy it contains \cite{schutz}. The current concordance model is the so-called $\Lambda$-Cold Dark Matter model ($\Lambda$CDM), which implements  the assumptions of homogeneity and isotropy to general relativity (which leads to the adoption of the Friedmann-Lema\^itre-Robertson-Walker, or FLRW, metric) and a mixture of pressureless dark matter and dark energy (the latter is assumed to be a cosmological constant).  The comparison with the observational data allows one to estimate the relative abundance of these two matter components and the age of the Universe \cite{cosmo}. It is fair to say that, for now, the $\Lambda$CDM cosmology remains as our best model of the Universe.\newline

There are many challenges to this framework that, however, remain to be addressed. For example, the theoretical estimate of the dark energy scale from quantum field theory and its inferred value in this framework differs by some 120 orders of magnitude \cite{wein}. There is also seemingly no good reason as well to why we (the observer performing the measurements) currently live in an epoch in which dark matter and dark energy abundances are of the same order of magnitude (the so-called ``coincidence problem'') \cite{velten}, and more importantly there is no \emph{direct} evidence of dark energy, which is considered the most urgent open question within the cosmological community \cite{survey}. Moreover, linear perturbation theory applied to this model cannot quite account for the formation of primordial black holes \cite{dolgov}. With these problems in mind, and maintaining general relativity as the correct theory of gravity, some inhomogeneous cosmological models have been proposed. Intuitive motivations behind them is that the aforementioned Copernican principle is just a philosophical and prejudicial  requirement \cite{phyl}, and that a detected almost-isotropic temperature distribution does not imply an almost-isotropic universe \cite{iso1,iso2}. For example, after the first detection of gravitational waves by the LIGO collaboration \cite{ligo}, current projects still cannot confirm whether their spectrum and distribution are indeed homogeneous and isotropic \cite{prc}. Therefore there are numerous reasons for considering inhomogeneous models other than Friedmann. In addition, there is no proof that small scale inhomogeneities (smaller than 100 Mpc \cite{millenium}) are irrelevant for the global evolution of the Universe as a whole \cite{comp1,comp2}.  \newline

The competition among inhomogeneous models themselves is fierce. It was first claimed that the Lema\^itre-Tolman-Bondi (LTB) model can account for the observed luminosity distance of the type Ia supernovae without the need of dark fluids but assuming that the Earth is placed inside a giant void  \cite{cliftonprl}, but later this model was ruled out by the kinematic Sunyaev-Zel'dovich effect \cite{bull}.  Recently it was proved that vacuum models based on regular lattices of black holes can support a negative deceleration parameter \cite{mio}, but they have not yet been tested against available data, that is, the amount of "accelerated expansion" may not be enough for describing the real Universe. On the other hand, inhomogeneous models displaying sharp profiles of the matter density, or of other cosmological parameters, have been derived as a proposed solution to the structure formation problem, but questions remain regarding whether they can be well-fitted with other datasets \cite{lim1,lim2,lim3}. Last but not least, the Stephani model has been proposed as a possibility of having a universe filled with a homogeneous energy density but supporting an inhomogeneous pressure \cite{st1,st2,st3,st4,st5,st6,st7,st8}. In this case the equation of state of the fluid permeating the universe is dynamical, meaning that it changes both in space and in time, thus allowing for different spacetime regions to be filled by different types of matter. \newline

The most widely adopted techniques for testing the cosmological model under investigation rely on astrophysical analyses and the studies of the distance modulus for the type Ia supernovae \cite{sup1,sup2,sup3}, the cosmic microwave background radiation \cite{cmb1,cmb2}, the baryon acoustic oscillations \cite{bao1,bao2}, the gamma ray bursts \cite{grb1,grb2,grb3} and the sum-distance rule \cite{bolejiko}. However, fundamental physics like thermodynamics can be exploited for confirming the validity of a certain cosmological model. For example, the  total entropy amount of the universe cannot decrease in time, in accordance to the second law of thermodynamics. Another requirement is that of ``cosmological holographic principle'': one should compare the amount of entropy inside the dynamical apparent horizon with the area of the horizon itself, see \cite{review} for a review. If one takes the holographic principle seriously, then the entropy of a physical cosmology should be bounded above by the horizon area. This principle has already been applied to the inhomogeneous Tolman-Lema\^itre-Bondi metric \cite{prl},  some (2+1)-dimensional cosmological models \cite{1+2}, and some dark energy models \cite{hol1,hol2,hol3},  with the one derived from the nonadditive Tsallis entropy being one example  \cite{TSA}, including  as well as to the cosmic microwave background radiation data \cite{hol4}. Recently a specific Stephani model was investigated along this line \cite{ong}. However, we found that this analysis contains both some inconsistent assumptions and technical mistakes which render its results doubtful, and a separate analysis should be performed {\it ab initio}. \newline

The aforementioned study, conducted by some of the present authors \cite{ong}, is unfortunately incorrect due to the following reasons.
The main result in \cite{ong} is that a certain Stephani metric based on a specific time evolution of the scale factor is compatible with the holographic requirement, but not with the type Ia supernova data. The analysis assumes that the type of matter permeating the universe is dust, in particular when estimating the numerical values of the free parameters entering the model from the dark matter abundance of the $\Lambda$CDM model. However, the equation of state supported by the cosmological fluid for the location of the observer considered there in the Stephani model is not dust but some matter that satisfies $p=-\frac{\rho}{3}$  as explained in section 2 of \cite{st7}. This exotic equation of state was first proposed in the context of straight cosmic strings for describing phases transitions in the early universe, but according to the current standard model of cosmology it does not play any role in the late time evolution \cite{string1,string2,string3}.  This completely undermines all the analysis therein. In addition, \cite{ong} also contains other computational mistakes, beyond this conceptual one, which essentially follow from missing speed-of-light factors and some wrong signs. For the sake of a clearer exposition, we summarize them in a table in the appendix to this paper, so that at least the interested reader can better understand the mathematics behind this project, but we will not comment on them any further because, as we mentioned, even after those corrections the model would still be physically meaningless because of a wrong physical assumptions. \emph{Therefore we need to re-open the investigation into the second law and holographic principle in Stephani cosmology.}\newline

The purpose of this paper is to establish a quantitative relationship between the abundance of inhomogeneities, the Hubble function, and the abundance of regular matter in a Stephani universe for accounting for the thermodynamical requirements. It is shown that the holographic principle allows us to set an upper bound on the inhomogeneity parameter and on the maximum spatial size that this type of universe can reach, and a lower bound on its entropy content, while the second law of thermodynamics allows us to restrict the range of validity of the values of the deceleration parameter. Our manuscript provides as well the opportunity of listing the mistakes which invalidate \cite{ong}. \newline

Our paper is organized as follows: in the next section we will exhibit the difficulties of constructing a simple explicit Stephani model with dust at the present epoch. In the third section we will argue that such explicitly constructed model is not required for concluding that Stephani cosmology is in agreement with both the holographic and increasing entropy requirements. Finally in the last section we conclude with some discussions about the viability of Stephani cosmology, and the virtue of our method of analysis. We provide at the end of this manuscript an appendix discussing some technical problems of the previous study \cite{ong}.

\section{Dust Modeling in a Stephani Model}

Consider the Stephani cosmological model given by the metric tensor
\beq
\label{eq1}
\d s^2=-\frac{c^2} {V^2(t,r)} \d t^2 +\frac{R^2(t)}{V^2(t,r)}[\d r^2+r^2   (\d\theta^2 +\sin^2 \theta \d \phi^2)],
\eeq
with
\begin{eqnarray}
V(r,t)&=&1-\frac{\beta}{c^2}R(t)r^2,\\
k(t)&=&-\frac{4\beta R(t)}{c^2}\,,
\end{eqnarray}
where $\beta$ is a free parameter of the model with dimension  $[\beta]= \text{km}^2/(\text{s}^2\text{Mpc})$. Moreover let $[R(t)]=$Mpc, $[r]=[V(t,r)]=0$, $[c]=$km/s, $[t]=$sMpc/km. Here $c$ is the speed of light and $G$ is the Newton constant of gravitation. In the limit $\beta \to 0$ the model reduces to a flat Friedmann universe. This model fulfills the condition $\frac{\text{d}}{\text{d}t} \left(\frac{k(t)}{R(t)}\right)=0$ and is more specifically known in the literature under the name ``Stephani model II' ' or D\c{a}browski model \cite{db1,db2,db3}. We prefer to focus on this class of the Stephani model, instead than on the Class I model, which is defined by the condition $\frac{\partial^2}{\partial t^2}\left(\frac{V(r,t)}{R(t)}  \right)  =0$,  because in this case it is possible to fully account for the strength of inhomogeneities just using one free parameter entering the metric (in fact these models are defined respectively through a first order and a second order differential equation) which will be shown to play a similar role than the cosmological constant in the Friedmann equation.  \newline

Now, let an over-dot denote derivative with respect to the cosmic time $t$, a semicolon a covariant derivative,  a subscript $0$ the present day values, and a subscript $e$ the quantities at the end of the radiation era. 
Consider the observer four-velocity $u^\mu=\frac{c}{V(t,r)} \delta^\mu_t$, $u_\mu u^\mu=-c^2$. The Hubble function is 
\beq
H(t):=\frac{u^\mu{}_{;\mu}}{3}=\frac{\dot R(t)}{R(t)}\,.
\eeq
Let
\begin{eqnarray}
q(r,t)&=&q(t)+\frac{\beta}{4}r^2R(t)(1+q(t)),\\
q(t)&:=&-\frac{\ddot R(t) R(t)}{\dot R^2(t)}
\end{eqnarray}
be the deceleration parameter. We stress that this model supports a homogeneous Hubble function (i.e. rate of expansion), but a space-dependent deceleration parameter as pointed out in \cite{stelmach}. Therefore the location of the observer is of crucial importance for detecting an accelerating or a decelerating expansion of the universe.  The Einstein equations are given by $G_{\mu\nu}=({8\pi G}/{c^4}) T_{\mu\nu}$, where $G_{\mu\nu}$ is the Einstein tensor and $T^\mu{}_{\nu}={\rm diag}[-\rho(t)c^2,\, p(t,r),\, p(t,r),\, p(t,r)]$ is the stress-energy tensor. In this model the energy density $\rho(t)$ is homogeneous, while the pressure $p(t,r)$ is inhomogeneous. The time-time component of the Einstein equation constitutes the generalized Friedmann equation
\beq
\label{gerf}
\left(\frac{\dot R}{R}\right)^2=\frac{4 \beta}{R}+\frac{8\pi G}{3 }\rho\,,
\eeq
which can be recast as the Gauss constraint 
\beq
\label{gauss}
1=\Omega_{\rm inhom}+\Omega_{\rm m},
\eeq
in which we have introduced the matter parameters
\beq
\label{cosmop}
\Omega_{\rm inhom}:=\frac{4 \beta}{R H^2}\,, \qquad \Omega_{\rm m}:=\frac{8\pi G}{3H^2 }\rho\,.
\eeq
In particular the former takes into account the effect of the inhomogeneities, and thus physical interpretation requires $\beta>0$, which implies that this model has a negative curvature $k(t)<0$. The latter comes with a subscript $m$ -- for ``matter'' -- it can account for a pressureless dust, a radiation fluid, or any other type of fluid. In the limit $\rho \to 0$ of a vacuum spacetime, the model can be easily integrated as
\beq
R(t)=\beta \left( t-t_0 +\sqrt{\frac{R_0}{\beta}}  \right)^2\,,
\eeq
and it is accelerating in the center of the universe because
\beq
\label{qcenter}
q(t, r=0)=-2\beta \frac{R}{\dot R^2}<0\,.
\eeq
The mix-rank spatial components of the  Einstein tensor  read as
\beq
G^r{}_{r}=G^\theta{}_{\theta}=G^\phi{}_{\phi}=\frac{2r^2 \beta R( rR{\ddot R}-\dot R^2+2\beta R)-c^2 (2R{\ddot R}+\dot R^2-8R\beta)}{R^2 c^4} \,.
\eeq
Having dust at the center of the universe at the present time requires $p(t_0, \, r=0)=0$, which implies $G^r{}_{r}(t_0, \, r=0)=G^\theta{}_{\theta}(t_0, \, r=0)=G^\phi{}_{\phi}(t_0, \, r=0)=0$, that is
\beq
2R_0{\ddot R}_0+\dot R_0^2-8R_0 \beta=0\,,
\eeq
which gives
\beq
\label{bb1}
\beta=\frac{1}{8}(1-2q_0)H_0^2 R_0\,.
\eeq
This in turn relates the amount of the inhomogeneities to the the deceleration parameter, the Hubble function and the present day curvature via
\beq
\label{beta0}
\beta=\frac{|H_0| c}{4}\sqrt{\left(q_0-\frac{1}{2}\right)k_0}\,.
\eeq
To have $\beta$ real (and in fact positive) requires $q_0 \leqslant {1}/{2}$. Let
\beq
\label{re}
r_e:=\int_0^{t_e}\frac{c}{R(t)}~\d t
\eeq
be the position of the emitter. Having a radiation fluid located at $r_e$ at time $t_e$ requires $p(t_e,\, r_e)=\frac{c^2 \rho(t_e)}{3}$. The two nontrivial components of the Bianchi identities $T^{\mu\nu}{}_{;\nu}=0$ read:
\begin{eqnarray}
&& \dot \rho R(c^2-R\beta r^2)+3\dot R(c^2 \rho +p)=0, \\
&& p'(R\beta r^2-c^2)-2\beta R r(\rho c^2 +p)=0\,,
\end{eqnarray}
where a prime denotes a derivative with respect to $r$. The latter can be integrated to
\beq
p(t,r)=(\beta R r^2 -c^2)F_1(t)-c^2 \rho\,,
\eeq
where $F_1(t)$ is an arbitrary integration constant. Then the former can be formally integrated as:
\beq
\rho(t)=3\int F_1(t) H(t) \d t +C\,,
\eeq
where $C$ is an arbitrary integration constant.
Choosing $F_1(t)=-\rho(t)$ we get
\beq
p(t,r)=-\beta R(t) r^2 \rho(t),
\eeq
which guarantees to have dust at the center of the Universe at the present time. However it is not possible to have radiation at $r_e$ at any time because $\beta>0$. Furthermore in this case the energy density dilutes as
\beq 
\label{consdust}
\rho(t)=\frac{\rho_0 R_0^3}{R^3(t)}\,,
\eeq
and the only field equation we are left to integrate is
\beq
R \dot R^2-4 R^2 \beta-\frac{8 \pi G \rho_0 R_0^3}{3}=0\,.
\eeq
Introducing a conformal time $\d t =R\d\eta$ we get the parametric solution 
\beq
R(t)=\beta^{-1/3}{\mathcal P}\left(\beta^{1/3}\eta +C_1, -\frac{8 \pi G \rho_0 R_0^3}{3 \beta^{1/3}}, 0 \right),
\eeq
in terms of the Weierstrass elliptic function $\mathcal P$, with $C_1$ being an integration constant.\newline

On the other hand, if we fix  $F_1(t)=\rho(t)$ instead, we would obtain
\beq
p(t,r)=(\beta R(t) r^2-2c^2) \rho(t),
\eeq
which is compatible with a radiation fluid as long as $\beta=\frac{2c^2}{R(t_e)r_e}$, but cannot describe dust at the center of this universe at any time. \newline

Following \cite{stelmach}, let 
\beq
R(t)=\alpha t^\gamma\,,
\eeq
with $[\gamma]=0$ and $[\alpha]=({\rm km/s})^\gamma{\rm Mpc}^{1-\gamma}$. Intuitively this is the simplest free-parameter model one can have. In fact in the time evolution of the scale factor we have one constant accounting for the power law behavior, and one constant for fixing the appropriate physical dimensions. With this choice we get
\beq
\frac{8\pi G}{c^4}p(t,r=0)=\frac{\alpha \gamma (2-3\gamma) +8\beta t^{2-\gamma}}{\alpha (ct)^2}\,,
\eeq
which can be interpreted as dust at the present time $ t_0$ if we impose
\beq
\label{eqa1}
\alpha=\frac{8\beta t_0^{2-\gamma}}{\gamma (3\gamma-2)}\,.
\eeq
Note that \cite{stelmach} writes $R(t) \sim t^{2/3}$; in this way $p(t, r=0) \sim t^{-2/3}$ which is dust for late times. However we want dust at current epoch $t_0$ instead. 
Moreover,
\begin{eqnarray}
H(t)&=&\frac{\gamma}{t}, \\
q(t, r=0)&=&\frac{1-\gamma}{\gamma}\,.
\end{eqnarray}
Therefore we see that an accelerating universe for an observer placed at the center of the configuration requires either $\gamma<0$ or $\gamma>1$. However a negative $\gamma$ implies a collapsing universe (since Hubble function is negative), while $\gamma>1$ causes the integral defining $r_e$ in (\ref{re}) to diverge at $t=0$. So it seems that all these conditions cannot be satisfied together.\newline

In fact this limitation is even worse. To see this, note that
\beq
H_0=\frac{\dot R_0}{R_0}=\frac{\alpha \gamma t_0^{\gamma-1}}{R_0}=\frac
{\alpha \gamma t_0^{\gamma-1}\Omega_{\rm inhom0}H_0^2}{4\beta}=\frac{2 t_0 \Omega_{\rm inhom0} H_0^2}{3\gamma-2}\,,
\eeq
where we have used (\ref{cosmop})  and (\ref{eqa1}), and therefore
\beq
\Omega_{\rm inhom0}=\frac{3\gamma-2}{2t_0 H_0}\,,
\eeq
or 
\beq
\gamma=\frac{2}{3}(\Omega_{\rm inhom0} t_0 H_0 +1)     \,.
\eeq
So there is the need of one more condition for fully fix all the free parameters $\alpha$, $\beta$, $\gamma$.
The previous formulas suggest two possibilities for fixing another parameter of the model:
\beq
\gamma=H_0 t_0 \,,
\eeq
or
\beq
\gamma=\frac{1}{1+q_0}\,.
\eeq
Thus we have the following possibilities summarized in Table 1, which show that it is not possible to reproduce the predicted present day value of dark energy abundance $\Omega_{\Lambda 0}$ (whose role in this model is played by the spatial inhomogeneities from (\ref{cosmop}) according to the interpretation $ \Omega_\Lambda \to \Omega_{\rm inhom}$) starting from the values of the Hubble function and of age of the universe provided by the analyses of the supernovae luminosity distance, the cosmic microwave background, the baryon acoustic oscillations and the gamma ray bursts discussed for example in \cite{sup1,sup2,sup3,cmb1,cmb2,bao1,bao2,grb1,grb2,grb3}.  

\begin{table*}
\label{tabella1}
\begin{center}
\begin{tabular}{ |l|l|l|l|l| }
\hline
$H_0$ in km s$^{-1}$Mpc${-1}$ &  $t_0$ in km$^{-1}$s Mpc & $\gamma$ & $q_0$ & $\Omega_{\rm inhom0}$ \\
  \hline
72 & 0.014 &1.008 &-0.008 & 0.51    \\
  \hline
71 & 0.014 &0.994 &0.006    &  0.49    \\
  \hline
72 & 0.013 &0.936 &0.068   &    0.43  \\
  \hline
\end{tabular}
\end{center}
\caption{ This table summarizes some possible combinations of various values for the cosmological parameters and the free parameters entering a two-parameter dust Stephani model. }
\end{table*}

\section{A general approach to the holographic principle in a Stephani cosmology}

In the previous section we have explained why it is not possible to construct a trivial time evolution of the scale factor for a Stephani cosmology fulfilling all the constraints we need for setting up a proper discussion of the holographic principle. Basic mathematics states that if we write
\beq
R(t)=\sum_{i=1}^{n} a_i t^{b_i},
\eeq
then we are likely to fulfill $n$ independent conditions by choosing properly the values of $n$ constants $a_i$ and $b_i$. However in this section we will show that it is possible to tackle the holographic principle and the second principle of thermodynamics in the Model II Stephani cosmology in a more elegant way with just  few basic assumptions on $R(t)$.\newline

Let $\tilde r=({R(t)}/{V(t,r)})\cdot r$ be the areal radius. The position of the dynamical apparent horizon $r_{\rm AH}$ can be obtained by imposing the condition $||\nabla \tilde r||^2=0$ \cite{prl, ong}. Thus we must solve the equation
\beq
(\beta R r_{\rm AH}^2 +c^2)^2 - c^2 r_{\rm AH}^2 {\dot R^2}=0\,,
\eeq
which admits four mathematical roots:
\beq
r_{\rm AH}= \pm \left( \frac{{\dot R} \pm \sqrt{{\dot R}^2-4\beta R}}{2\beta R}\right)c  \,,
\eeq
where all the possible sign combinations must be considered. For understanding which is the physically relevant solution we start by rewriting the four roots in terms of the cosmologically meaningful parameters as:
\beq
r_{\rm AH}= \pm \frac{H c}{2\beta}(1\pm \sqrt{\Omega_{\rm m}})  \,.
\eeq
We will from now on assume this universe to be expanding, i.e. to have a positive Hubble function. The relevant solution for $r_{\rm AH}$ must be positive, and this already eliminates the possibilities $(-~+)$ and $(-~-)$, where for the latter case we used the constraint that the matter parameter has to be smaller than unity. Then we note that the solution must be {\it small} because we are assuming that the observer is located at the center of the universe $r=0$, and we must take into account that this model of the Stephani universe exhibits a spatial section consisting of two disconnected sheets (because its curvature is negative)  of which only the one containing the spatial origin, .i.e. the location of the observer, can be of physical interest  \cite{st6}.  The "branch point"  $r_B(t)$,  which separates the two spatial sheets is such that $V(t,r_B)=0$, which provides the condition
\beq
r_B(t)=2|k(t)|^{-1/2}\,.
\eeq
Thus we must impose $r_{\rm AH} <r_B$, i.e. that the dynamical apparent horizon must belong to the so-called {\it near sheet}. This would eliminate the choice $(+~+)$, leaving us with the only well-defined solution
\beq
\label{hor}
r_{\rm AH} (t)= \frac{H c}{2\beta}(1- \sqrt{\Omega_{\rm m}})\,.
\eeq
We need $r_{\rm AH} (t_0)<r_B(t_0)$, and $r_{\rm AH} (t_e)<r_B(t_e)$, which are true as long as
\begin{eqnarray}
&& 1- \sqrt{\Omega_{\rm m0}}< \sqrt{\frac{1}{2}-q_0}~,\\
&& 1- \sqrt{\Omega_{\rm m }(t_e)}< \sqrt{\frac{1}{2}-q(t_e)}\,.
\end{eqnarray}\newline

The area of the dynamical apparent horizon is
\begin{eqnarray}
A(t) &=& 4 \pi {\tilde r}_{\rm AH}^2 \,= \, \pi \left[ \frac{4 c R H(1-\sqrt{\Omega_{\rm m}})}{4 \beta -RH^2 (1-\sqrt{\Omega_{\rm m}})^2} \right]^2 \nonumber\\
&=& \pi \left[ \frac{4 c^3 (-k) H(1-\sqrt{\Omega_{\rm m}})}{16 \beta^2 +k c^2 H^2 (1-\sqrt{\Omega_{\rm m}})^2} \right]^2\,.
\end{eqnarray}
The time evolution of the "matter" entropy is
\beq
S_{\rm m} (t)= {\mathcal C} r_{\rm AH}^3\,,
\eeq
where we have introduced an overall constant (note that having radiation or another type of fluid affects only the value of this constant and not the previous formula)
\beq
{\mathcal C}= \frac{4  \kappa_B^4}{135 }\cdot\left( \frac{\pi T}{\hbar} c^2 \right)^3\cdot \left(\frac{R(t_e)}{V(t_e,r_e)}  \right)^3\,,
\eeq
where
$T$ is the temperature, $\kappa_B$ is the Boltzmann constant, and $\hbar$ is the reduced Planck constant. 
Now for establishing the validity of the holographic principle we must see under which conditions we can satisfy the inequality
\beq
\label{holo1}
\frac{S_{\rm m}(t)}{A(t)}>1\,,
\eeq
and to check the second law of thermodynamics, to see if
\beq
\frac {\text{d}S_{\rm m}(t)}{\text{d}t}>0\,.
\eeq
\emph{These two inequalities mean that the entropy inside the region bounded by the dynamical apparent horizon must be larger than the area of this region, and that the entropy must be an increasing function of the time.}

\subsection{Holographic Principle }

\begin{figure}
\begin{center}
\includegraphics[scale=0.8]{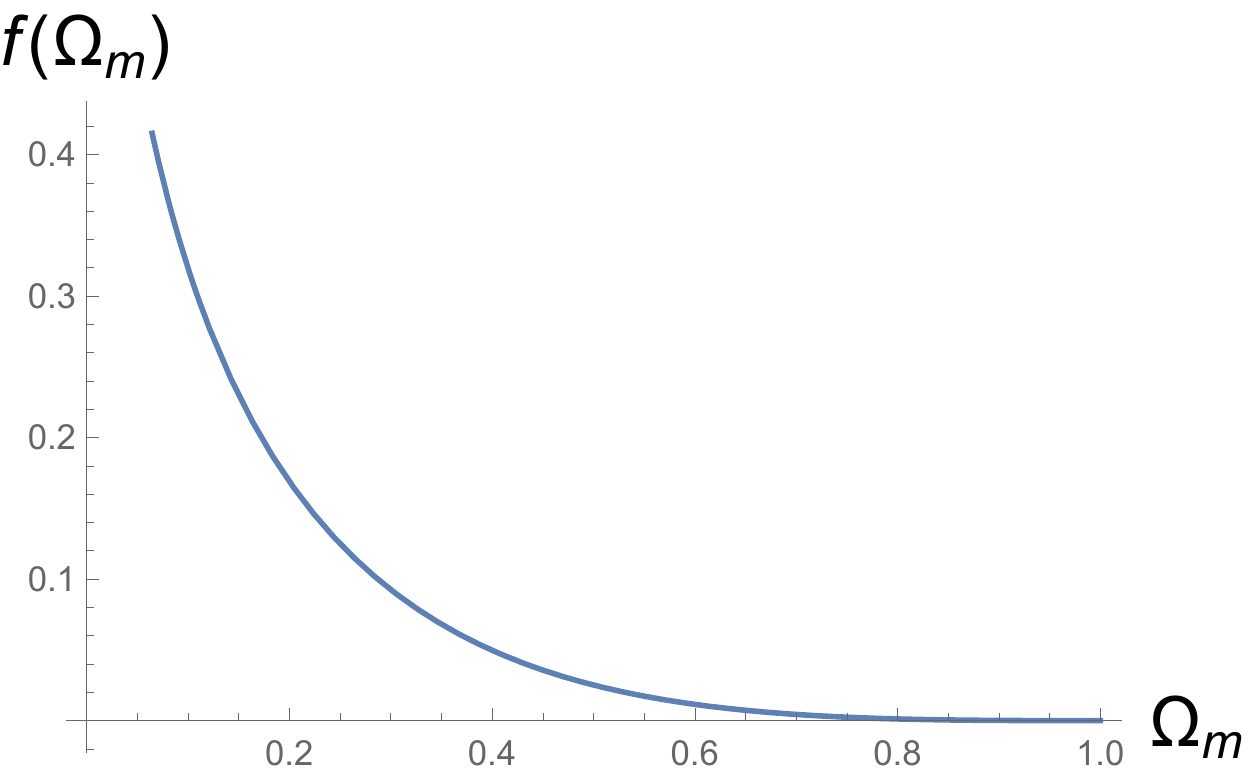}
\end{center}
\caption{ The figure confirms that the function  $f(\Omega_{\rm m})= 3 \Omega_{\rm m}+1-3 \Omega^{1/2}_{\rm m}- \Omega^{3/2}_{\rm m}$ is positive-definite in the interval of physical interest. }
\label{fig1}
\end{figure}

Inequality (\ref{holo1}) can be rewritten as
\begin{eqnarray}
\frac{{\mathcal C}}{4\pi}\cdot \frac{ r_{\rm AH}^3}{\tilde r_{\rm AH}^2}=\frac{{\mathcal C}}{4\pi}\cdot\frac{r_{\rm AH} V^2(t,r_{\rm AH})}{R^2(t)}=\frac{{\mathcal C}r_{\rm AH}}{4\pi} \left(\frac{c^2 -\beta R r^2_{\rm AH}}{c^2 R} \right)^2>1
\end{eqnarray}
Taking into account that a multiplication by the same positive factor on both sides does not change the sense of an inequality, using the location of the dynamical apparent horizon from (\ref{hor}) and replacing the metric scale factor in terms of the matter-energy density from (\ref{cosmop}) as
\beq
\label{size}
R=\frac{4\beta}{H^2(1-\Omega_{\rm m} )},
\eeq
we can recast the holographic principle requirement into:
\beq
{\mathcal C} H^5 c \Omega_{\rm m}  [ 3 \Omega_{\rm m}+1-3 \Omega^{1/2}_{\rm m}- \Omega^{3/2}_{\rm m}]-32\pi \beta^3 >0\,,
\eeq
in which we note that the function $f(\Omega_{\rm m})= 3 \Omega_{\rm m}+1-3 \Omega^{1/2}_{\rm m}- \Omega^{3/2}_{\rm m}$ is positive-definite in the interval $0\leq  \Omega_{\rm m}\leq 1$ as shown in figure (\ref{fig1}). Therefore the holographic principle is trivially satisfied in a flat Friedmann universe for which $\beta \to 0$, and it sets an upper bound to the abundance of inhomogeneities in the Stephani model as
\beq
\beta <\left(\frac{{\mathcal C} H^5 c \Omega_{\rm m}  [ 3 \Omega_{\rm m}+1-3 \Omega^{1/2}_{\rm m}- \Omega^{3/2}_{\rm m}]}{32 \pi}    \right)^{\frac{1}{3}}=\left(\frac{{\mathcal C} H^5 c \Omega_{\rm m} }{4 \pi}    \right)^{\frac{1}{3}}\frac{ 1-\sqrt{\Omega_{\rm m}}}{2}\,.
\eeq
Therefore the holographic principle sets an upper bound to the maximum size the Stephani universe can evolve to via Eq.(\ref{size}). Furthermore, using Eq.(\ref{beta0}) it is possible to set a lower bound for the present day entropy amount as
\beq
{\mathcal C}>\frac{c^2 \pi ((2q_0-1)k_0)^{3/2}}{4\sqrt{2} H^2_0 \Omega_{\rm m 0} [ 1- \sqrt{\Omega_{\rm m 0}}]^3}
\eeq

\subsection{Second Law of Thermodynamics}

\begin{figure}
\begin{center}
\includegraphics[scale=0.8]{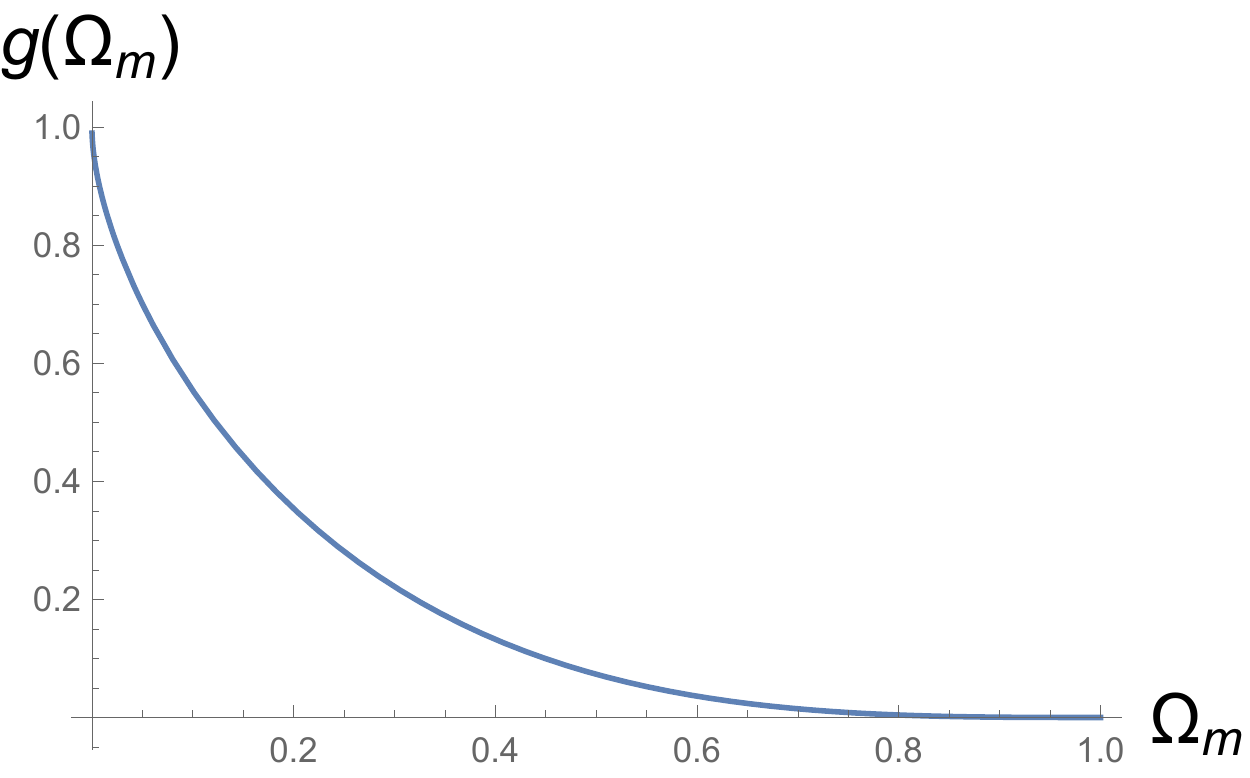}
\end{center}
\caption{ The figure confirms that the function  $g(\Omega_{\rm m})=2\Omega^{3/2}_{\rm m}-\Omega^{5/2}_{\rm m}-\Omega^{1/2}_{\rm m}+(1-\Omega_{\rm m})^2$ is positive-definite in the interval of physical interest. }
\label{fig2}
\end{figure}

\begin{figure}
\begin{center}
\includegraphics[scale=0.8]{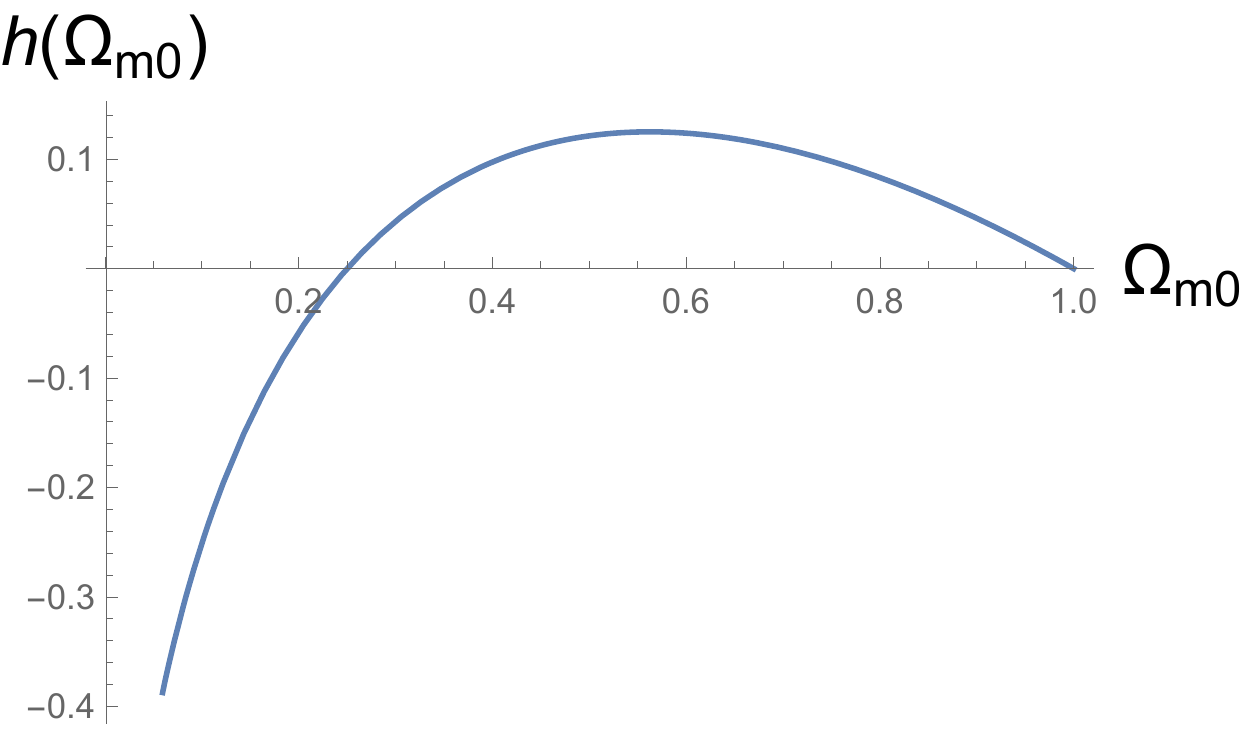}
\end{center}
\caption{ The figure shows the behavior of the function the function  $h(\Omega_{\rm m 0})=3 \sqrt{\Omega_{\rm m 0}}-2 \Omega_{\rm m 0}- 1$  in the interval of physical interest. }
\label{fig3}
\end{figure}

To check the validity of the second law of thermodynamics, we must see whether $\dot r_{\rm AH}>0$. The following results may be handy. From the time derivative of the generalized Friedmann equation (\ref{gerf}) we get
\beq
\label{dotH}
\dot H=-\frac{2\beta}{R}-\frac{4 \pi G\rho_0 R_0^3}{R^3}<0\,
\eeq
where we used the law of energy conservation for dust (\ref{consdust}). Moreover, 
\beq
\dot \Omega_{\rm m}=-\dot \Omega_{\rm inhom}=\frac{4\beta}{RH} \left[1+\frac{2\dot H}{H^2} \right]= \frac{(1-\Omega_{\rm m})H}{32 \beta^3} \left[32 \beta^3 \Omega_{\rm m} - 4 \pi G\rho_0 R_0^3 H^4 (1-\Omega_{\rm m})^3  \right]  \,,
\eeq
where in the last step we used (\ref{size}) and (\ref{dotH}). Therefore
\begin{eqnarray}
\dot r_{\rm AH}&=&\frac{1}{2\beta}\left[ \dot H (1-\sqrt{\Omega_{\rm m}})c-\frac{H}{2\sqrt{\Omega_{\rm m}}}\cdot \dot \Omega_{\rm m}c \right]\\
&=& \frac{H^2 (1-\Omega_{\rm m})c}{32 \sqrt{\Omega_{\rm m}} \beta^4}\Big[ \pi G\rho_0 R_0^3H^4(2\Omega^{3/2}_{\rm m}-\Omega^{5/2}_{\rm m}-\Omega^{1/2}_{\rm m}+(1-\Omega_{\rm m})^2) -8 \Omega^{1/2}_{\rm m} \beta^3   \Big]\nonumber\\
&=&  \frac{H^2 (1-\Omega_{\rm m})c}{32 \sqrt{\Omega_{\rm m}} \beta^4}\Big[ \pi G\rho_0 R_0^3H^4 (1-\Omega_{\rm m})^2 (1-\sqrt{\Omega_{\rm m}}) -8 \Omega^{1/2}_{\rm m} \beta^3   \Big]\nonumber\\
&=&  \frac{H^2 (1-\Omega_{\rm m})c}{4 \sqrt{\Omega_{\rm m}} \beta}\Big[\frac{3 \Omega_{\rm m 0}}{H^4_0 (1- \Omega_{\rm m 0})^3}H^4 (1-\Omega_{\rm m})^2 (1-\sqrt{\Omega_{\rm m}})- \Omega^{1/2}_{\rm m}    \Big]\nonumber
\end{eqnarray}
where in the last step we used Eq.(\ref{size}) and Eq.(\ref{cosmop}) specified at the present time. We note that the function $g(\Omega_{\rm m})=2\Omega^{3/2}_{\rm m}-\Omega^{5/2}_{\rm m}-\Omega^{1/2}_{\rm m}+(1-\Omega_{\rm m})^2= (1-\Omega_{\rm m})^2 (1-\sqrt{\Omega_{\rm m}})$ is positive-definite in the region of physical interest as shown in figure (\ref{fig2}). Imposing the validity of the second law of thermodynamics at the present time $t_0$ we get the condition:
\beq
3 \sqrt{\Omega_{\rm m 0}} (1-\Omega_{\rm m 0})^2 (1-\sqrt{\Omega_{\rm m 0}}) - (1- \Omega_{\rm m 0})^3>0\,,
\eeq
which can be simplified as
\beq
3 \sqrt{\Omega_{\rm m 0}}-2 \Omega_{\rm m 0}- 1>0\,  .
\eeq
Therefore figure (\ref{fig3}) sets a limit to the energy amount that the spatial inhomogeneities contribute to as $\Omega_{\rm inhom 0}<0.75$. Using (\ref{bb1}),  (\ref{cosmop}) and the latter upper bound we finally get an estimate of the deceleration parameter at the present time as
\beq
q_0 \sim -0.25\,.
\eeq
It is interesting to note that the second law of thermodynamics suggests that this model of the universe undergoes an accelerated expansion at the present time even though there are no dark fluids driving its evolution, contrary to the case of the Friedmann model commonly adopted in the standard cosmological modeling, but thanks to the role played by the spatial inhomogeneities. Moreover,  this procedure allows an alternative estimate of the deceleration parameter which does not rely on the supernovae data. The fact that its absolute value is smaller than the one inferred in a Friedmann framework should not come as a surprise because other inhomogeneous cosmological models predict similar properties, the timescape cosmology just being one example \cite{asta2}, and for this reason it does not rule out the validity of the Stephani model but it must be interpreted as one of its genuine physical feature.

\section{Conclusions}

Estimates of the values of the cosmological parameters, like the energy-matter abundance and the expansion history of the universe, are usually done within the Friedmann paradigm. The experimental points are likely to change when a different geometrical model is assumed (see for example the figures in \cite{cliftonprl} about how the Union 2.1 data points should be re-derived for the Tolman-Lema\^itre-Bondi model even before testing the model). In fact many observation techniques are based on reconstructing the travel of light rays which will differ in spacetimes with different curvature. Since these modifications are non-trivial, forcing a non-Friedmann model to be Friedmann-like in its rate of expansion, its deceleration, its present day amount of entropy, its age, its matter abundance, etc, as done in the previous analysis of the holographic principle in a Stephani universe \cite{ong} is highly unnatural and conceptually unmotivated. In fact particles are subject to completely different types of gravitational potentials during their motion, because in the former case they experience a curvature which is different in each point along their path while in the latter case only an {\it average} effect of the spacetime curvature is taken into account. A different gravitational potential will first of all affect the relationship between redshift and cosmic time and may contribute as well with a drift term to former. For example, \cite{cliftond} already quantified possible quantitative deviations to the abundance of regular matter due to the presence of inhomogeneities.  Only recently, some model-independent techniques for testing non-isotropic and non-homogeneous universe have been developed \cite{asta1,asta2}. Therefore  studies of cosmological models without referencing to Friedmann are scarce. In this paper we have proposed a route for testing the holographic principle and the second law of thermodynamics in a Model II Stephani cosmology, which does not require us to force the model to be Friedmann-like. As a sanity check we recover the already known results for the former in the limit of a flat homogeneous universe even without the problem of the tension between small or large scale estimates of the Hubble parameter \cite{tension}. Since our analysis of the holographic principle does not rely on any assumption on the matter content of the universe, our results can be applied both to late time cosmology or to the early universe whose evolutions are likely to be driven by different type of fluids, dust in the former or stringy scalar fields in the latter. 

The model considered here is shear-free. Thus we performed our analysis in terms of the rate of expansion and of the matter parameters. However the shear may play a role in the early universe even though its present day amount is negligible \cite{shear}, and its presence should be accounted for when studying thermodynamics during the inflationary epoch. This provides us the opportunity to repeat our analysis in the class of spacetime metrics \cite{sc1,sc2,sc3,sc4,sc5} which are shearing and supported by a stiff fluid. Since the latter is equivalent to a massless scalar field, we can analyze how it affects the entropic requirements for an early universe as well in a future study.

\begin{acknowledgments}
Y.C.O.  thanks NNSFC (grant No.11705162) and the Natural Science Foundation of Jiangsu Province
(No.BK20170479) for funding support. D.G. acknowledges support from China Postdoctoral Science Foundation (grant No.2019M661944).

\end{acknowledgments}

\appendix  \section{List of corrections to \cite{ong}}

This appendix summarizes a list of numerical corrections to \cite{ong}. Note that a dust model  is assumed, which contradicts the actual equation of state for regular matter adopted. Even though after the modifications proposed in this appendix \cite{ong} remains conceptually wrong, we think that they can help the reader in becoming familiar with the mathematical manipulations involved in this project.

\begin{center}
\begin{longtable}{ |l|l|l| }
\hline
  Position in the manuscript & Published statement& Correct statement  \\
  \hline
  Line below Eq. (9)     &  $ C(t)=A \cdot R(t) $ & $C(t)=\frac{ A}{ R(t)} $   \\
  \hline
Eq. (10), 4th line      & $\gamma=\pm \sqrt{4\beta \eta +1 }$    &  $\gamma=\pm \sqrt{4\beta \eta +c^2 }$  \\
  \hline
Eq. (21)      & $d_H (t_0) = \frac{\beta t_0^2 + \sqrt{4\beta\eta+1}t_0 + \eta}{1-\frac{\beta}{c^2}(\beta t_0^2 + \sqrt{4\beta\eta+1}t_0+\eta)r_0^2}\cdot r_0$   & $d_H (t_0) = \frac{\beta t_0^2 + \sqrt{4\beta\eta+c^2}t_0 + \eta}{1-\frac{\beta}{c^2}(\beta t_0^2 + \sqrt{4\beta\eta+c^2}t_0+\eta)r_0^2}\cdot r_0$   \\
  \hline
 Eq. (22)      &  $r_0:=\int_0^{t_0} \frac{c}{\beta t^2 + \sqrt{4\beta\eta+1}t + \eta} \text{d}t' \approx 3.6052 \times 10^5$   & \makecell[l]{ $r_0:=\int_0^{t_0} \frac{c}{\beta t^2 + \sqrt{4\beta\eta+c^2}t + \eta} \text{d}t'$ \\ $ \approx 1.201774235 $ }   \\
  \hline
  Eq. (25)    &  $k=...\simeq -2.8573 \times 10^{-11}$   &   $k=...\simeq  -2.571487349$  \\
  \hline 
Eq. (26)      &   $r_B(t_0) = 3.7416 \times 10^5$  &  $r_B(t_0) = 1.247204875$   \\
  \hline     
Sentence below Eq. (26) & \makecell[l]{ \noindent The fact that $k$ is extremely small is, \\ by itself, 
a good thing, since the \\ observed spatial curvature \\ of the Universe is
close to being 
flat.}     & \makecell[l]{The sentence should be removed \\because it
 does not hold  \\ anymore}    \\
  \hline
Three line below Eq. (26)      &   $r_0\approx 3.6052 \times 10^5$  &  $r_0 \approx 1.201774235$   \\
  \hline 
Eq. (27)&  $r_{\text{AH}}^2 \dot{R}^2(t) + (\frac{k(t)}{2}r_{\text{AH}}^2 - \frac{k^2(t)}{16}r_{\text{AH}}^4 -1) =0$   &  \makecell[l]{ $\Big[(-r_{\text{AH}}^3 k(t)-4 r_{\text{AH}})\dot R(t)+\dot k(t) r_{\text{AH}}^3 R(t)  $ \\ $+c r_{\text{AH}}^2 k(t)-4 c\Big] \cdot \Big[(r_{\text{AH}}^3 k(t)+4 r_{\text{AH}}) \dot R(t)$\\$-  \dot k(t) r_{\text{AH}}^3 R(t)+c r_{\text{AH}}^2 k(t)-4 c \Big] =0$}   \\
  \hline
Eq. (28)       & \makecell[l]{ $r_{\text{AH}}^2 (2\beta t + \sqrt{4\beta\eta+1})^2 - (  \frac{2\beta(\beta t^2 + \sqrt{4\beta\eta+1}t+\eta)}{c^2}r_{\text{AH}}^2$\\$-\frac{\beta^2(\beta t^2 + \sqrt{4\beta\eta+1}t+\eta)^2}{c^4}r_{\text{AH}}^4-1)=0$}   &     \makecell[l]{$ \Big[\beta (\beta t^2+\sqrt{4\beta\eta+c^2}) t+\eta) r^2$ \\$+2 (\beta t+1/2 \sqrt{4\beta\eta+c^2}) ) c r +c^2 \Big]$\\ $\cdot \Big[\beta (\beta t^2+\sqrt{4\beta\eta+c^2}) t+\eta) r^2$\\$-2(\beta t+1/2 \sqrt{4\beta\eta+c^2})) c r+c^2\Big]=0$  }  \\
  \hline
 Eq. (30)     &   $r_1\approx 0.5291$, $r_2\approx 2.6456 \times 10^{11}$  & $r_1 \approx 0.6920 $,  $r_2 \approx  2.2476$    \\
  \hline
Eq. (32)      & $\tilde r_1\approx 0.0139$ Mpc $\sim 10^{22}$ cm     &  \makecell[l]{ $\tilde r_{1,2}=\pm  7874.324654$ Mpc \\ $= \pm 0.2429747178 \times 10^{29}$ cm  }  \\
  \hline
Sentence below Eq. (32)      &  \makecell[l]{ This value is much smaller than the \\
 particle horizon $10^{29 }$ cm $\approx 32407.7929$ Mpc.}    &  \makecell[l]{ Only the positive value \\associated to $r_1$ is physical. \\ This value is of the same order of \\ magnitude of the particle horizon \\ $10^{29 }$ cm $\approx 32407.7929$ Mpc.}   \\
  \hline
Sentence below Eq. (32)  & \makecell[l]{Therefore, unlike the 
flat FLRW  universe \\ case, in which the size of the particle \\ horizon is comparable to that of the \\ apparent horizon, the apparent
horizon in \\ this  Stephani model is much smaller \\ than its  particle horizon. }    &   \makecell[l]{Therefore, like the 
flat FLRW  universe \\ case, in which the size of the particle \\ horizon is comparable to that of the \\ apparent horizon, the apparent \\
horizon in  this  Stephani  model \\ is close  to its  particle horizon. }     \\
  \hline
Sentence below Eq. (42)      &     &\makecell[l]{ where \\ $r_e=0.7059733793 \times 10^{-6}$ }   \\
  \hline
\makecell[l]{ First line of the footnote \\ on page 7 }       &  \makecell[l]{ The large difference between  the values \\ of the two roots is due to large numbers \\ involved in  the coefficients, such as $c$. }   &\makecell[l]{ This sentence should be removed\\ because it does not hold anymore. }   \\
  \hline
Eq. (44)      &  $r_{\rm AH } = r_1(t_e) \approx 0.8305 $  & \makecell[l]{ $ r_\text{AH}=r_1(t_e)=0.90742071810$. \\ Note that it remains true that \\ $r_1(t_e) > r_1(t_0)$. }   \\
  \hline
Fig. (1) on page 9     &    & See below  \\
  \hline
Eq. (45)      &  $S_m \approx 1.0742 \times 10^{ 58}$ J/K   & \makecell[l]{ $S_m =  0.964643 \times 10^{15 } $ J/K for $r_1$} \\
  \hline
Sentence below Eq. (45)      &   $S_m \sim 10^{ 81} $ & \makecell[l]{ these correspond to $S_m \sim 10^{41}$, \\ and $S_m \sim 10^{39}$ respectively} \\
  \hline
 Eq. (47)     & $\rho= \,{\frac {3 }{8\pi\,G \left( \beta\,{t}^{2
}+\gamma\,t+\eta \right) ^{2}}}>0$    & $\rho= \,{\frac {3 c^2}{8\pi\,G \left( \beta\,{t}^{2
}+\gamma\,t+\eta \right) ^{2}}}>0$    \\
  \hline
 Eq. (48)     & $\rho+3P/c^2=-\,{\frac {\,{3\beta}{r}^{2}/c^2}{4\pi\,G \left( \beta
\,{t}^{2}+\gamma t+\eta \right)}} > 0$    & $\rho+3P/c^2=-\,{\frac {\,{3\beta}{r}^{2}}{4\pi\,G \left( \beta
\,{t}^{2}+\gamma t+\eta \right)}} > 0$    \\
  \hline
 Eq. (49)     & $\rho+P/c^2= \,{\frac {1-\,{\beta}{r}^{2}(\beta\,{t}^{2
}+\gamma\,t+\eta)/c^2}{4\pi\,G \left( \beta
\,{t}^{2}+\gamma t+\eta \right) ^{2}}} > 0$    & $\rho+P/c^2= \,{\frac {c^2-\,{\beta}{r}^{2}(\beta\,{t}^{2
}+\gamma\,t+\eta)}{4\pi\,G \left( \beta
\,{t}^{2}+\gamma t+\eta \right) ^{2}}} > 0$    \\
  \hline
  Eq. (50)    &   $\rho-P/c^2=\,{\frac {2+\,{\beta}{r}^{2}(\beta\,{t}^{2
}+\gamma\,t+\eta)/c^2}{4\pi\,G \left( \beta\,{
t}^{2}+\gamma t+\eta \right) ^{2}}} > 0$  & $\rho-P/c^2=\,{\frac {2c^2+\,{\beta}{r}^{2}(\beta\,{t}^{2
}+\gamma\,t+\eta)}{4\pi\,G \left( \beta\,{
t}^{2}+\gamma t+\eta \right) ^{2}}} > 0$    \\
  \hline
Eq. (54)      &  $\sqrt{4\beta\eta+1}\approx 0.04916\beta$   &  $\sqrt{4\beta\eta+c^2}\approx 0.04916\beta$   \\
  \hline
 Eq. (55)     &  $72 = \frac{2\beta t_0 + \sqrt{4\beta\eta+1}}{\beta t_0^2 + \sqrt{4\beta\eta+1}+\eta}$   & $72 = \frac{2\beta t_0 + \sqrt{4\beta\eta+c^2}}{\beta t_0^2 + t_0\sqrt{4\beta\eta+c^2}+\eta}$    \\
  \hline
 Eq. (56)     & $\beta \approx 24.4925$, $\eta \approx 0.004591$     & \makecell[l]{$\beta \approx 7.3473 \times 10^{6}$  km$^2$/(s$^2$ $\cdot$ Mpc), \\
$\eta \approx 1377.161824 $ Mpc}   \\
  \hline
 Eq. (59)     & \makecell[l]{ $m=M+25+5\log_{10}\left[cz\left(\frac{\beta t^2-\gamma t+\eta}{2\beta t-\gamma}\right)\right] $\\$+1.086z\left[1+2\beta \frac{(\beta t^2-\gamma t+\eta)}{(2\beta t-\gamma)^2}\right]$}     & \makecell[l]{$m=M+25+5\log_{10}\left[cz\left(\frac{\beta t^2+\gamma t+\eta}{2\beta t+\gamma}\right)\right] $\\$+1.086z\left[1+4\beta \frac{(\beta t^2+\gamma t+\eta)}{(2\beta t+\gamma)^2}\right]$ \\ Note that the correct formula \\ appears in \cite{st7} and not in \cite{db1}}   \\
  \hline
 Eq. (61)     &      & \makecell[l]{This equation requires $\omega=-\frac{1}{3}$\\ and not $\omega=0$ as instead used in the \\ paper for comparing model vs. \\supernova data.}   \\
  \hline
 Eq. (61)     &      & \makecell[l]{Note that at this time, the most \\ recent  supernovae catalog is \\ called {\it pantheon} \\ \cite{sup3} and not Union 2.1}   \\
  \hline
\caption{ This table summarizes a list of technical mistakes in  \href{https://link.springer.com/article/10.1140/epjc/s10052-018-5866-1}{EPJC { \bf 78} (2018) 405}.  }
\end{longtable}
\label{chi2}
\end{center}


To summarize (ignoring the conceptual mistake about the type of matter component and consequently the physical interpretation of the model): the Stephani model considered in \cite{ong} can account for the supernova data, as contrary written there (one can easily see this as well from the deceleration parameter $q=-2\beta<0$ and the independent analysis in \cite{st7,db1,stelmach}), and satisfies the cosmological holographic principle for an appropriate amount of inhomogeneities.

{}


\begin{thebibliography}{99}


\bibitem{schutz}
Bernard Schutz,  ``A First Course in General Relativity ",  (Cambridge University Press, Cambridge, 2009).

\bibitem{cosmo}
George Francis Rayner Ellis, and Henk van Elst, ``Cosmological models (Carg\'ese lectures 1998)",
NATO Adv. Study Inst. Ser. C. Math. Phys. Sci.  \textbf{1}  (1999), \href{https://arxiv.org/abs/gr-qc/9812046}{[arXiv:gr-qc/9812046]}. 

\bibitem{wein}
Steven Weinberg, ``The cosmological constant problem", 
{\hypersetup{urlcolor=vividviolet}\href{https://journals.aps.org/rmp/abstract/10.1103/RevModPhys.61.1}{Rev. Mod. Phys. \textbf{61}  (1989)  1}}.

\bibitem{velten}
Hermano E.S. Velten,  Rodrigo von Marttens,  and Winfried Zimdahll,  ``Aspects of the cosmological  {\it coincidence problem}",
{\hypersetup{urlcolor=vividviolet}\href{https://link.springer.com/article/10.1140%2Fepjc%2Fs10052-014-3160-4}{Eur. Phys. J. C {\bf 74} (2014)  3160}}, \href{https://arxiv.org/abs/1410.2509}{ [arXiv:astro-ph.CO/1410.2509]}.

\bibitem{survey}
Krzysztof Bolejko, Mikołaj Korzy\'nski, ``Inhomogeneous cosmology and backreaction: current status and future prospects",
{\hypersetup{urlcolor=vividviolet}\href{https://www.worldscientific.com/doi/abs/10.1142/S0218271817300117}{Int. J. Mod. Phys. D  \textbf{26}  (2017)  1730011}}, \href{https://arxiv.org/abs/1612.08222}{[arXiv:gr-qc/1612.08222]}.

\bibitem{dolgov}
Alexander D. Dolgov, ``Beasts in Lambda-CDM Zoo",
{\hypersetup{urlcolor=vividviolet}\href{https://link.springer.com/article/10.1134%2FS1063778817050076}{Physics of Atomic Nuclei  \textbf{80}  (2017)   5}}, \href{https://arxiv.org/abs/1605.06749}{[arXiv:astro-ph.CO/1605.06749]}.

\bibitem{phyl}
George Francis Rayner Ellis, ``The Homogeneity of the Universe", 
{\hypersetup{urlcolor=vividviolet}\href{https://link.springer.com/article/10.1007%2FBF00759562}{Gen. Rel. Grav.  \textbf{11}  (1979)  4}}.

\bibitem{iso1}
Ulf S. Nilsson, Claes Uggla, John Wainwright, and Woei Chet Lim,  ``An almost isotropic cosmic microwave temperature does not imply an almost isotropic universe",  
{\hypersetup{urlcolor=vividviolet}\href{https://iopscience.iop.org/article/10.1086/312209/meta}{Astrophys.J.  {\bf 522}  (1999)  L1}}, \href{https://arxiv.org/abs/astro-ph/9904252}{[arXiv:astro-ph/9904252]}.

\bibitem{iso2}
Woei Chet  Lim, Ulf S. Nilsson,  and John Wainwright,  ``Anisotropic universes with isotropic cosmic microwave background radiation",
{\hypersetup{urlcolor=vividviolet}\href{https://iopscience.iop.org/article/10.1088/0264-9381/18/24/317/meta}{Class.  Quantum  Grav.  {\bf 18}  (2001)   5583}}, \href{https://arxiv.org/abs/gr-qc/9912001}{[arXiv:gr-qc/9912001]}.

\bibitem{ligo}
The LIGO Scientific Collaboration,  and the Virgo Collaboration,  ``Observation of Gravitational Waves from a Binary Black Hole Merger",
{\hypersetup{urlcolor=vividviolet}\href{https://journals.aps.org/prl/abstract/10.1103/PhysRevLett.116.061102}{Phys. Rev. Lett. \textbf{116}  (2016) 061102}}, \href{https://arxiv.org/abs/1602.03837}{[arXiv:gr-qc/1602.03837]}.

\bibitem{prc}
Jun Luo, Li-Sheng Chen, Hui-Zong Duan, Yun-Gui Gong, Shoucun Hu, Jianghui Ji, Qi Liu, Jianwei Mei, Vadim Milyukov, Mikhail Sazhin, Cheng-Gang Shao, Viktor T. Toth, Hai-Bo Tu, Yamin Wang, Yan Wang, Hsien-Chi Yeh, Ming-Sheng Zhan, Yonghe Zhang, Vladimir Zharov, and Ze-Bing Zhou, ``TianQin: a space-borne gravitational wave detector",  
{\hypersetup{urlcolor=vividviolet}\href{https://iopscience.iop.org/article/10.1088/0264-9381/33/3/035010/pdf}{Class. Quantum Grav.  \textbf{33}  (2016)  035010}}, \href{https://arxiv.org/abs/1512.02076}{[arXiv:astro-ph.IM/1512.02076]}.

\bibitem{millenium}
Volker Springel, Simon D. M. White, Adrian Jenkins, Carlos S. Frenk, Naoki Yoshida, Liang Gao, Julio Navarro, Robert Thacker, Darren Croton, John Helly, John A. Peacock, Shaun Cole, Peter Thomas, Hugh Couchman, August Evrard, Joerg Colberg,  and Frazer Pearce, ``Simulating the joint evolution of quasars, galaxies and their large-scale distribution",
{\hypersetup{urlcolor=vividviolet}\href{https://www.nature.com/articles/nature03597}{Nature  \textbf{435}  (2005)  629}}, \href{https://arxiv.org/abs/astro-ph/0504097}{[arXiv:astro-ph/0504097]}.

\bibitem{comp1}
Chris Clarkson, George Ellis, Julien Larena and Obinna Umeh, ``Does the growth of structure affect our dynamical models of the universe? The averaging, backreaction, and fitting problems in cosmology",
{\hypersetup{urlcolor=vividviolet}\href{http://iopscience.iop.org/article/10.1088/0034-4885/74/11/112901/meta}{Rept. Prog. Phys.  \textbf{74}  (2011)   112901}}, \href{https://arxiv.org/abs/1109.2314}{[arXiv:astro-ph.CO/1109.2314]}.

\bibitem{comp2}
Thomas Buchert, Mauro Carfora, G.F.R. Ellis, Edward W. Kolb, Malcolm MacCallum, Jan J. Ostrowski, Sysky R\"as\"anen,  Boudewijn F. Roukema, Lars Andersson, Alan Coley, and David L. Wiltshire, ``Is there proof that backreaction of inhomogeneities is irrelevant in cosmology?",
{\hypersetup{urlcolor=vividviolet}\href{http://iopscience.iop.org/article/10.1088/0264-9381/32/21/215021/meta}{Class. Quantum Grav.  \textbf{32}  (2015)   215021}}, \href{https://arxiv.org/abs/1505.07800}{[arXiv:gr-qc/1505.07800]}.

\bibitem{cliftonprl}
Timothy Clifton, Pedro G. Ferreira, and  Kate Land,  ``Living in a Void: Testing the Copernican Principle with Distant Supernovae", 
{\hypersetup{urlcolor=vividviolet}\href{https://journals.aps.org/prl/abstract/10.1103/PhysRevLett.101.131302}{Phys. Rev. Lett.  \textbf{101}  (2008)   131302}}, \href{https://arxiv.org/abs/0807.1443}{[arXiv:astro-ph/0807.1443]}.

\bibitem{bull}
Philip Bull, Timothy Clifton, and  Pedro G. Ferreira, ``The kSZ effect as a test of general radial inhomogeneity in LTB cosmology",
{\hypersetup{urlcolor=vividviolet}\href{https://journals.aps.org/prd/abstract/10.1103/PhysRevD.85.024002}{Phys. Rev. D  \textbf{85}  (2012)  024002 }}, \href{https://arxiv.org/abs/1108.2222}{[arXiv:astro-ph.CO/1108.2222]}.

\bibitem{mio}
Timothy Clifton, Daniele Gregoris, Kjell Rosquist, and  Reza Tavakol, ``Exact Evolution of Discrete Relativistic Cosmological Models",
{\hypersetup{urlcolor=vividviolet}\href{https://iopscience.iop.org/article/10.1088/1475-7516/2013/11/010}{JCAP  \textbf{11}  (2013)  010}}, \href{https://arxiv.org/abs/1309.2876}{[arXiv:gr-qc/1309.2876]}.

\bibitem{lim1}
Alan Coley, and Woei Chet Lim, ``Spikes and matter inhomogeneities in massless scalar field models",
{\hypersetup{urlcolor=vividviolet}\href{http://iopscience.iop.org/article/10.1088/0264-9381/33/1/015009/meta}{Class. Quantum Grav.  \textbf{33}  (2016)   015009}}, \href{https://arxiv.org/abs/1511.07095}{[arXiv:gr-qc/1511.07095]}.

\bibitem{lim2}
Alan Coley, Daniele Gregoris, Woei Chet Lim,  ``On the first $G_1$ stiff fluid spike solution in General Relativity", 
{\hypersetup{urlcolor=vividviolet}\href{https://iopscience.iop.org/article/10.1088/0264-9381/33/21/215010/meta}{Class. Quantum Grav.  \textbf{33}  (2016)  215010}}, \href{https://arxiv.org/abs/1606.07177}{[arXiv:gr-qc/1606.07177]}.

\bibitem{lim3}
Daniele Gregoris, Woei Chet Lim, Alan Coley,  ``Stiff fluid spike solutions from Bianchi type V seed solutions",
{\hypersetup{urlcolor=vividviolet}\href{https://iopscience.iop.org/article/10.1088/1361-6382/aa924f/meta}{Class. Quantum Grav.  \textbf{34}  (2017)  235013}}, \href{https://arxiv.org/abs/1705.02747}{[arXiv:gr-qc/1705.02747]}.

\bibitem{st1}
Hans Stephani,  ``\"Uber L\"osungen der Einsteinschen Feldgleichungen, die sich in einen f\"unfdimensionalen flachen Raum einbetten lassen",
{\hypersetup{urlcolor=vividviolet}\href{https://link.springer.com/article/10.1007%2FBF01645757}{Commun. Math. Phys.  {\bf 4}  (1967) 137}}.

\bibitem{st2}
Carles Bona, and Bartolom\'e Coll,  ``On the Stephani Universes",   {\hypersetup{urlcolor=vividviolet}\href{https://link.springer.com/article/10.1007%2FBF00759188}{Gen. Relat. Gravit {\bf 20}  (1988) 297}}. 

\bibitem{st3}
Hans Stephani, Dietrich Kramer, Malcolm MacCallum, Cornelius Hoenselaers, and Eduard Herlt, `` Exact Solutions of Einstein's
Field Equations",  (Cambridge Monographs on Mathematical Physics, Cambridge, 2009).

\bibitem{st4}
 Max Wyman, ``Equations of State for Radially Symmetric Distributions of Matter",  {\hypersetup{urlcolor=vividviolet}\href{https://journals.aps.org/pr/abstract/10.1103/PhysRev.70.396}{Phys. Rev. {\bf 70}  (1946) 396}}.      

\bibitem{st5}
Paul Kustaanheimo, and Bertil Qvist,  ``A Note on Some General Solutions of the Einstein Field Equations in a Spherically
Symmetric World",  {\hypersetup{urlcolor=vividviolet}\href{https://link.springer.com/article/10.1023/A:1018882511219}{Gen. Rel. Grav. {\bf 30} (1998) 663}}. 

\bibitem{st6}
Andrzej Krasi\'nski,  ``On the Global Geometry of the Stephani Universe", {\hypersetup{urlcolor=vividviolet}\href{https://link.springer.com/article/10.1007/BF00759044}{Gen. Rel. Grav. {\bf 15} (1982) 673}}. 

\bibitem{st7}
 Mariusz P. D\c{a}browski, and Martin A. Hendry, ``The Hubble Diagram of Type Ia Supernovae in Non-Uniform Pressure
Universes", {\hypersetup{urlcolor=vividviolet}\href{https://iopscience.iop.org/article/10.1086/305546/meta}{Astrophys. J. {\bf 498}  (1998) 67}}, \href{https://arxiv.org/abs/astro-ph/9704123}{[arXiv:astro-ph/9704123]}. 

\bibitem{st8}
Adam Balcerzak, Mariusz P. D\c{a}browski, Tomasz Denkiewicz, David Polarski, and Denis Puy,  ``A Critical Assessment of Some
Inhomogeneous Pressure Stephani Models", 
{\hypersetup{urlcolor=vividviolet}\href{https://journals.aps.org/prd/abstract/10.1103/PhysRevD.91.083506}{Phys. Rev. D {\bf 91}  (2015) 083506}}, \href{https://arxiv.org/abs/1409.1523}{[arXiv:gr-qc/1409.1523]}. 

\bibitem{sup1}
Adam G. Riess, Alexei V. Filippenko, Peter Challis, Alejandro Clocchiattia, Alan Diercks, Peter M. Garnavich, Ron L. Gilliland, Craig J. Hogan, Saurabh Jha, Robert P. Kirshner, B. Leibundgut, M. M. Phillips, David Reiss, Brian P. Schmidt, Robert A. Schommer, R. Chris Smith, J. Spyromilio, Christopher Stubbs, Nicholas B. Suntzeff, and John Tonry, ``Observational Evidence from Supernovae for an Accelerating Universe and a Cosmological Constant",
{\hypersetup{urlcolor=vividviolet}\href{https://iopscience.iop.org/article/10.1086/300499/meta}{Astron. J.  \textbf{116}  (1998)  1009}}, \href{https://arxiv.org/abs/astro-ph/9805201}{[arXiv:astro-ph/9805201]}.

\bibitem{sup2}
Saul Perlmutter, Greg Aldering, Gerson Goldhaber, Robert Knop, Peter Nugent, Patricia G. Castro, Susana Deustua, Sebastien Fabbro, Ariel Goobar, Donald E. Groom, Isobel  M. Hook, Alex G. Kim, Mattew Kim, Julia C. Lee, Nelson J. Nunes, Reynald Pain, Carl R. Pennypacker, Robert Quimby, Chris Lidman, Richard S. Ellis, Mike Irwin, Richard G. McMahon, Pilar Ruiz-Lapuente, Nicholas Walton, Bradley Schaefer, Brian J. Boyle, Alexei V. Filippenko, Thomas  Matheson, Andrew S. Fruchter, Nino Panagia, Heidi J.M. Newberg, and  Warrick J. Couch (The Supernova Cosmology Project), ``Measurements of Omega and Lambda from 42 High-Redshift Supernovae", 
{\hypersetup{urlcolor=vividviolet}\href{https://iopscience.iop.org/article/10.1086/307221/meta}{Astrophys. J.  \textbf{517}  (1999)  565}}, \href{https://arxiv.org/abs/astro-ph/9812133}{[arXiv:astro-ph/9812133]}.

\bibitem{sup3}
Daniel M. Scolnic, David O. Jones, Armin Rest, Yen-Chen Pan, Ryan Chornock, Ryan J. Foley, Mark  E. Huber, Richard  Kessler, Gautham Narayan, Adam G. Riess, Stephen Rodney, Edo Berger, Dillion J. Brout, Peter J. Challis, Maria Drout, Douglas Finkbeiner, Ragnhild Lunnan, Robert P. Kirshner, Nathan E. Sanders, Edward Schlafly, Stephen Smartt, Christopher W. Stubbs, John Tonry, William Michael. Wood-Vasey, Michael Foley, John Hand, Erik Johnson, William S. Burgett, Kenneth C. Chambers, P. W. Draper, Klaus-Werner. Hodapp, Norbert Kaiser, Rolf-Peter Kudritzki, Eugene A. Magnier, Nigel Metcalfe, Fabio Bresolin, E. Gall, Rubina Kotak, Matt  McCrum, and Kester W. Smithl,  ``The Complete Light-curve Sample of Spectroscopically Confirmed Type Ia Supernovae from Pan-STARRS1 and Cosmological Constraints from The Combined Pantheon Sample",
{\hypersetup{urlcolor=vividviolet}\href{https://iopscience.iop.org/article/10.3847/1538-4357/aab9bb/meta}{Astrophys. J.  \textbf{859}  (2018)  2}}, \href{https://arxiv.org/abs/1710.00845}{[arXiv:astro-ph.CO/1710.00845]}.

\bibitem{cmb1}
Eiichiro Komatsu, Kester M. Smith, Joanna Dunkley, Charles L. Bennett, Benjamin Mark Gold, Gary Hinshaw, N. Jarosik, Delbert John Larson, Michael R. Nolta, Lyman A. Page, David  N. Spergel, Mark Halpern, Robert S. Hill, Alan Kogut, Michele Limon, Stephen S. Meyer, Nils Odegard, Gregory S. Tucker, Janet L. Weiland, Edward Wollack, and Edward L. Wright, ``Seven-Year Wilkinson Microwave Anisotropy Probe (WMAP) Observations: Cosmological Interpretation",
{\hypersetup{urlcolor=vividviolet}\href{http://iopscience.iop.org/article/10.1088/0067-0049/192/2/18/meta}{Astrophys. J. Suppl.  \textbf{192}  (2011)  18}}, \href{https://arxiv.org/abs/1001.4538}{[arXiv:astro-ph.CO/1001.4538]}.

\bibitem{cmb2}
Ryan Keisler, Christian L. Reichardt, K. A. Aird, Bredford A. Benson, Lindsey E. Bleem, John E. Carlstrom, Clarence Leeder Chang,  Hsiao-Mei Cho, Timothy M. Crawford, Abigail T. Crites, Tijmen de Haan, Mattew A. Dobbs, Jonhatan Dudley, Elizabeth M. George, Nils W. Halverson, Gilbert P. Holder, William  L. Holzapfel, Stephen Hoover, Z. Hou, J. D. Hrubes, Marshall Joy, Lloyd Knox, Aaron T. Lee, Erik M. Leitch, Martin  Lueker, Daniel  Luong-Van, Jeff J. McMahon, Jared Mehl, Steffen S. Meyer, Marius Millea, Joseph J. Mohr, Thomas E. Montroy, Tyler Natoli, Stephen Padin, Thomas Plagge, Clement Pryke, John E. Ruhl, Kathryn Kelly Schaffer, Laurie Shaw, Erik Shirokoff, Helmuth G. Spieler, Z. Staniszewski, Antony A. Stark, Kyle Story, Alexander van Engelen, Keith Vanderlinde, Joaquin D. Vieira, Robert Williamson, and Olivier Zahn, ``A Measurement of the Damping Tail of the Cosmic Microwave Background Power Spectrum with the South Pole Telescope",
{\hypersetup{urlcolor=vividviolet}\href{http://iopscience.iop.org/article/10.1088/0004-637X/743/1/28/meta}{ Astrophys. J.  \textbf{743}  (2011)  1}}, \href{https://arxiv.org/abs/1105.3182}{[arXiv:astro-ph.CO/1105.3182]}.

\bibitem{bao1}
Daniel J. Eisenstein, Idit Zehavi, David W. Hogg, Roman Scoccimarro, Michael R. Blanton, Robert C. Nichol, Ryan Scranton, Hee-Jong Seo, Max Tegmark, Zheng Zheng, Scott F. Anderson, Jim Annis, Neta Bahcall, Jon Brinkmann, Scott Burles, Francisco J. Castander, Andrew Connolly, Istvan Csabai, Mamoru Doi, Masataka Fukugita, Joshua A. Frieman, Karl Glazebrook, James E. Gunn, John S. Hendry, Gregory Hennessy, Zeljko Ivezi\'e, Stephen Kent, Gillian R. Knapp, Huan Lin, Yeong-Shang Loh, Robert H. Lupton, Bruce Margon, Timothy A. McKay, Avery Meiksin, Jeffery A. Munn, Adrian Pope, Michael W. Richmond, David Schlegel, Donald P. Schneider, Kazuhiro Shimasaku, Christopher Stoughton, Michael A. Strauss, Mark SubbaRao, Alexander S. Szalay, Istv\'an Szapudi, Douglas L. Tucker, Brian Yanny, and Donald G. York,  ``Detection of the Baryon Acoustic Peak in the Large-Scale Correlation Function of SDSS Luminous Red Galaxies",
{\hypersetup{urlcolor=vividviolet}\href{http://iopscience.iop.org/article/10.1086/466512/meta}{Astrophys. J. \textbf{633}  (2005)  560}}, \href{https://arxiv.org/abs/astro-ph/0501171}{[arXiv:astro-ph/0501171]}.

\bibitem{bao2}
Shaun Cole, Will J. Percival, John A. Peacock, Peder Norberg, Carlton M. Baugh, Carlos S. Frenk, Ivan Baldry, Joss Bland-Hawthorn, Terry Bridges, Russell Cannon,  Matthew Colless,  Chris Collins, Warrick Couch,  Nicholas J. G. Cross,  Gavin Dalton,  Vincent R. Eke,  Roberto de Propris,  Simon P. Driver,  George Efstathiou, Richard S. Ellis,  Karl Glazebrook,  Carole Jackson,  Adrian Jenkins,  Ofer Lahav,  Ian Lewis,  Stuart Lumsden,  Steve Maddox,  Darren Madgwick,  Bruce A. Peterson,  Will Sutherland, and Keith Taylor (The 2dFGRS Team),  ``The 2dF Galaxy Redshift Survey: Power-spectrum analysis of the final dataset and cosmological implications",
{\hypersetup{urlcolor=vividviolet}\href{https://academic.oup.com/mnras/article/362/2/505/1017047}{Mon. Not. Roy. Astron. Soc. \textbf{362}  (2005)  505}}, \href{https://arxiv.org/abs/astro-ph/0501174}{[arXiv:astro-ph/0501174]}. 

\bibitem{grb1}
Marek Demianski, Ester Piedipalumbo, Disha Sawant, and Lorenzo Amati, ``Cosmology with gamma-ray bursts: I. The Hubble diagram through the calibrated E$_{\rm p,i }$-E$_{\rm iso}$ correlation",  
{\hypersetup{urlcolor=vividviolet}\href{https://www.aanda.org/articles/aa/abs/2017/02/aa28909-16/aa28909-16.html}{A. and A.  \textbf{598}  (2017)   A112}}, \href{https://arxiv.org/abs/1610.00854}{[arXiv:astro-ph.CO/1610.00854]}.

\bibitem{grb2}
Hannachi Zitouni, Nidhal Guessoum, and Jamil Walid Azzam, ``Determination of Cosmological Parameters from Gamma Ray Burst Characteristics and Afterglow Correlations",  
{\hypersetup{urlcolor=vividviolet}\href{https://link.springer.com/article/10.1007%2Fs10509-016-2969-8}{Astrophys. Space Sci. \textbf{361}  (2016)  383}}, \href{https://arxiv.org/abs/1612.06369}{[arXiv:astro-ph.HE/1612.06369]}.

\bibitem{grb3}
Lorenzo Amati, and Massimo Della Valle,  ``Measuring cosmological parameters with Gamma-Ray Bursts",
{\hypersetup{urlcolor=vividviolet}\href{https://www.worldscientific.com/doi/abs/10.1142/S0218271813300280}{Int. J. Mod. Phys. D \textbf{22}  (2013)  1330028}}, \href{https://arxiv.org/abs/1310.3141}{[arXiv:astro-ph.CO/1310.3141]}.

\bibitem{bolejiko}
Syksy R\"as\"anen, Krzysztof Bolejko, and Alexis Finoguenov,  ``New Test of the Friedmann-Lemaitre-Robertson-Walker Metric Using the Distance Sum Rule",
{\hypersetup{urlcolor=vividviolet}\href{https://journals.aps.org/prl/abstract/10.1103/PhysRevLett.115.101301}{Phys. Rev. Lett. {\bf 115} (2015) 101301 }}, \href{https://arxiv.org/abs/1412.4976}{[arXiv:astro-ph.CO/1412.4976]}.

\bibitem{review}
Raphael Bousso, ``The holographic Principle", 
{\hypersetup{urlcolor=vividviolet}\href{https://journals.aps.org/rmp/abstract/10.1103/RevModPhys.74.825}{Rev. Mod. Phys. \textbf{74}  (2002) 825}}, \href{https://arxiv.org/abs/hep-th/0203101}{[arXiv:hep-th/0203101]}.

\bibitem{prl}
Bin Wang, Elcio Abdalla, and Takeshi Osada, ``Entropy and Holography Constraints for Inhomogeneous Universes'', {\hypersetup{urlcolor=vividviolet}\href{https://journals.aps.org/prl/abstract/10.1103/PhysRevLett.85.5507}{Phys. Rev. Lett. \textbf{85} (2000) 5507}}, \href{https://arxiv.org/abs/astro-ph/0006395}{[arXiv:astro-ph/0006395]}.

\bibitem{1+2}
Bin Wang, and Elcio Abdalla, ``Holography In (2+1)-Dimensional Cosmological Models'', {\hypersetup{urlcolor=vividviolet}\href{http://www.sciencedirect.com/science/article/pii/S0370269399011223}{Phys. Lett. B \textbf{466} (1999) 122}}, \href{https://arxiv.org/abs/hep-th/9910185}{[arXiv:hep-th/9910185]}.

\bibitem{hol1}
Bin Wang, Elcio Abdalla, and Ru-Keng Su, ``Constraints on the Dark Energy from Holography'', {\hypersetup{urlcolor=vividviolet}\href{https://www.sciencedirect.com/science/article/pii/S037026930500242X}{Phys. Lett. B \textbf{611} (2005) 21}}, \href{https://arxiv.org/abs/hep-th/0404057v3}{[arXiv:hep-th/0404057]}.

\bibitem{hol2}
Yungui Gong, Bin Wang, and Yuan-Zhong Zhang, ``The Holographic Dark Energy Reexamined'', {\hypersetup{urlcolor=vividviolet}\href{https://journals.aps.org/prd/abstract/10.1103/PhysRevD.72.043510}{Phys. Rev. D \textbf{72} (2005) 043510}}, \href{https://arxiv.org/abs/hep-th/0412218v3}{[arXiv:hep-th/0412218]}.

\bibitem{hol3}
Jianyong Shen, Bin Wang, Elcio Abdalla, Ru-Keng Su, ``Constraints on the Dark Energy from the Holographic Connection to the Small  $l$  CMB Suppression",
{\hypersetup{urlcolor=vividviolet}\href{https://www.sciencedirect.com/science/article/pii/S037026930500119X?via%3Dihub}{Phys. Lett. B  {\bf 609}  (2005) 200}}, \href{https://arxiv.org/abs/hep-th/0412227}{[arXiv:hep-th/0412227]}. 

\bibitem{TSA}
Rocco D'Agostino, ``Holographic dark energy from nonadditive entropy: cosmological perturbations and observational constraints", 
{\hypersetup{urlcolor=vividviolet}\href{https://journals.aps.org/prd/abstract/10.1103/PhysRevD.99.103524}{Phys. Rev. D  {\bf 99} 103524 (2019)}}, \href{https://arxiv.org/abs/1903.03836}{[arXiv:gr-qc/1903.03836]}. 


\bibitem{hol4}
Zhuo-Yi Huang, Bin Wang, Elcio Abdalla, Ru-Keng Su, ``Holographic Explanation of Wide-Angle Power Correlation Suppression in the Cosmic Microwave Background Radiation'', 	{\hypersetup{urlcolor=vividviolet}\href{https://iopscience.iop.org/1475-7516/2006/05/013/}{JCAP \textbf{0605} (2006) 013}}, \href{https://arxiv.org/abs/hep-th/0501059}{[arXiv:hep-th/0501059]}.

\bibitem{ong}
Yen Chin Ong, S. Sedigheh Hashemi, Rui An, and Bin Wang ``Stephani cosmology: entropically viable but observationally challenged",
{\hypersetup{urlcolor=vividviolet}\href{https://link.springer.com/article/10.1140%2Fepjc%2Fs10052-018-5866-1}{Eur. Phys. J. C \textbf{78}  (2018) 405}}, \href{https://arxiv.org/abs/1712.02297}{[arXiv:gr-qc/1712.02297]}.

\bibitem{string1}
Alexander Vilenkin,  ``Cosmic strings and domain walls", 
{\hypersetup{urlcolor=vividviolet}\href{https://www.sciencedirect.com/science/article/abs/pii/037015738590033X?via%3Dihub}{Phys. Rep.  \textbf{121}  (1985) 263}}.

\bibitem{string2}
Edward W. Kolb, and Michael S. Turner, ``The early universe", Front. Phys. {\bf 69} (1990).

\bibitem{string3}
Alexander Vilenkin, and E. Paul Shellard, ``Cosmic strings and other topological defects",   (Cambridge Monographs on Mathematical Physics, Cambridge, 1994).

\bibitem{db1}
Mariusz P. D\c{a}browski, ``A Redshift-Magnitude Relation for Non-Uniform Pressure Universes", 
{\hypersetup{urlcolor=vividviolet}\href{http://articles.adsabs.harvard.edu/cgi-bin/nph-iarticle_query?1995ApJ...447...43D&amp;data_type=PDF_HIGH&amp;whole_paper=YES&amp;type=PRINTER&amp;filetype=.pdf}{Astrophys. J. {\bf 447}  (1995)
43}}, \href{https://arxiv.org/abs/0902.2899}{[arXiv:astro-ph.CO/0902.2899]}.

\bibitem{db2}
Mariusz P. D\c{a}browski,  ``Isometric Embedding of the Spherically Symmetric Stephani Universe: Some Explicit Examples",
{\hypersetup{urlcolor=vividviolet}\href{https://aip.scitation.org/doi/10.1063/1.530166}{Journ. Math. Phys. {\bf 34}  (1993) 1447}}.

\bibitem{db3}
Richard K. Barrett, and Chris A. Clarkson,  ``Undermining the Cosmological Principle: Almost Isotropic Observations in
Inhomogeneous Cosmologies",
{\hypersetup{urlcolor=vividviolet}\href{https://iopscience.iop.org/article/10.1088/0264-9381/17/24/308/meta}{Class. Quant. Grav. {\bf 17}  (2000) 5047}}, \href{https://arxiv.org/abs/astro-ph/9911235}{[arXiv:astro-ph/9911235]}.

\bibitem{stelmach}
Izabela Jakacka, and Jerzy Stelmach,  ``Non-homogeneity-driven Universe acceleration",
{\hypersetup{urlcolor=vividviolet}\href{https://iopscience.iop.org/article/10.1088/0264-9381/18/14/303/meta}{Class. Quant. Grav. {\bf 18}  (2001) 2643}}, \href{https://arxiv.org/abs/0802.2284}{[arXiv:astro-ph/0802.2284]}.

\bibitem{suss}
Willy Fischler, and Leonard Susskind, ``Holography and Cosmology",    {\hypersetup{urlcolor=vividviolet}\href{}{}} \href{https://arxiv.org/abs/hep-th/9806039}{[arXiv:hep-th/9806039]}.   

\bibitem{asta2}
Asta Heinesen, Chris Blake, Yong-Zhuang Li, and David L. Wiltshire, ``Baryon acoustic oscillation
methods for generic curvature: Application to the SDSS-III Baryon Oscillation Spectroscopic
Survey", 
{\hypersetup{urlcolor=vividviolet}\href{https://iopscience.iop.org/article/10.1088/1475-7516/2019/03/003/meta}{JCAP {\bf 03}  (2019) 003}}, \href{https://arxiv.org/abs/1811.11963}{[arXiv:astro-ph.CO/1811.11963]}.

\bibitem{cliftond}
Timothy Clifton, and Pedro. Ferreira, ``Errors in Estimating 
 $\Omega_\Lambda$ due to the Fluid Approximation", 
{\hypersetup{urlcolor=vividviolet}\href{https://iopscience.iop.org/article/10.1088/1475-7516/2009/10/026/meta}{JCAP
{0910} (2009) 26}}, \href{https://arxiv.org/abs/0908.4488}{[arXiv:astro-ph.CO/0908.4488]}.

\bibitem{asta1}
Lawrence H. Dam, Asta Heinesen, and David L. Wiltshire, ``Apparent cosmic acceleration from type
Ia supernovae",
{\hypersetup{urlcolor=vividviolet}\href{https://academic.oup.com/mnras/article-abstract/472/1/835/4002695?redirectedFrom=fulltext}{Mon. Not. Roy. Astron. Soc. {\bf 472}  (2017) 835}}, \href{https://arxiv.org/abs/1706.07236}{[arXiv:astro-ph.CO/1706.07236]}.

\bibitem{tension}
Licia Verde, Pavlos Protopapas, and Raul Jimenez,  ``Planck and the Local Universe: Quantifying the Tension",
{\hypersetup{urlcolor=vividviolet}\href{https://www.sciencedirect.com/science/article/pii/S2212686413000319?via%3Dihub}{ Phys. Dark
Univ. {\bf 2} (2013) 166}}, \href{https://arxiv.org/abs/1306.6766}{[arXiv:astro-ph.CO/1306.6766]}.

\bibitem{shear}
Mahmoud Hashim ,  ``Bulk and Shear Viscosity in the Early Universe Cosmology: Dissipative QCD Cosmology", (Lap Lambert Academic Publishing, 2011).

\bibitem{sc1}
Clement Leibovitz,   ``Time-Dependent Solutions of Einstein's Equations", 
{\hypersetup{urlcolor=vividviolet}\href{https://journals.aps.org/prd/abstract/10.1103/PhysRevD.4.2949}{Phys. Rev. D \textbf{4} (1971) 2949}}.

\bibitem{sc2}
Kayll Lake,  ``Remark concerning spherically symmetric nonstatic solutions to the Einstein equations in the comoving frame",
{\hypersetup{urlcolor=vividviolet}\href{https://link.springer.com/article/10.1007%2FBF00759164}{Gen. Rel. Grav. \textbf{15} (1983) 357}}.

\bibitem{sc3}
Norbert van den Bergh, and  Patrick Wils,  ``Exact solutions for nonstatic perfect fluid spheres with shear and an equation of state",
{\hypersetup{urlcolor=vividviolet}\href{https://link.springer.com/article/10.1007%2FBF00760245}{Gen. Rel. Grav. \textbf{17} (1985)  223}}.

\bibitem{sc4}
Barry Collins, and J.M. Lang,  ``A class of self-similar perfect-fluid spacetimes, and a generalisation", 
{\hypersetup{urlcolor=vividviolet}\href{http://iopscience.iop.org/article/10.1088/0264-9381/4/1/009/meta}{Class. Quantum Grav. \textbf{4} (1987) 61}}.

\bibitem{sc5}
Sunil D. Maharaj, Roy Maartens, and Monoj Maharaj, ``A note on a class of spherically symmetric solutions",
{\hypersetup{urlcolor=vividviolet}\href{https://link.springer.com/article/10.1007%2FBF02874342}{Nuovo Cim. B \textbf{108} (1993) 1}}.




\end{thebibliography}
\end{document}